\documentclass[aps,twocolumn,amsmath,amssymb,prl]{revtex4-1}
\usepackage{graphicx}

\usepackage{upgreek}
\newcommand{\muT}{\upmu {\rm T}}

\begin{document}

\title{A magic magnetic field to measure the neutron electric dipole moment}

\author{Guillaume Pignol}
 \email{guillaume.pignol@lpsc.in2p3.fr}
  \affiliation{%
Univ. Grenoble Alpes, CNRS, Grenoble INP, LPSC-IN2P3, 38000 Grenoble, France
}

\date{\today}

\begin{abstract}
New sources of CP violation beyond the Standard Model of particle physics could be revealed in the laboratory by measuring a non-zero electric dipole moment (EDM) of a spin 1/2 particle such as the neutron. 
Despite the great sensitivity  attained after 60 years of developments, the result of the experiments is still compatible with zero. 
Still, new experiments have a high discovery  potential since they probe new physics at the multi-TeV scale, beyond the reach of direct searches at colliders. 
Progress in precision on the neutron EDM is limited by a systematic effect arising from the relativistic motional field $\vec{E} \times \vec{v} / c^2$ experienced by the particles moving in the measurement chamber in combination with the residual magnetic gradients. 
This effect would normally forbid a significant increase of the size of the chamber, sadly hindering the increase of  neutron statistics. 
We propose a new measurement concept to evade this limitation in a room-temperature experiment employing a mercury co-magnetometer. 
It consists ajusting the static magnetic field $B_0$ to a ``magic'' value  which cancels the false EDM of the mercury. 
The magic setting is $7.2\,\muT$ for a big cylindrical double-chamber of diameter $100$~cm.

%This is the abstract, which is supposed to tell in a catchy but precise manner what the article is about. Here it is. At the very beginning of the article. You are reading the abstract right now. The abstract (here) is usually about four to six lines long. Let's write a fake abstract that long in order to see what it should look like. You are almost done reading the abstract, but not quite. That's it, you are done now. 
\end{abstract}

\pacs{Valid PACS appear here}
\maketitle

In 1950, Purcell and Ramsey \cite{PurcellRamsey1950} proposed to measure the Electric Dipole Moment (EDM) of the neutron, although it is predicted to be zero if one assumes the invariance of the laws of physics under parity. 
Since then, experiments to measure the neutron EDM improved in sensitivity by no less than six orders of magnitude, and yet the most recent measurement \cite{Pendlebury2015} is still compatible with zero:
\begin{equation}
\label{EDMlimit}
d_n = (-0.21 \pm 1.82) \times 10^{-26} \ e \, {\rm cm}. 
\end{equation}
Experiments measuring the EDM of other systems (see \cite{Jungmann2013} for a recent review) are also restlessly reporting zero with continuously increased precision since decades. 

In fact, the existence of a non-zero permanent EDM for the neutron, or any non-degenerate system of spin 1/2, would not only violate parity $\mathrm{P}$ but also the invariance under time reversal $\mathrm{T}$. 
In turn, according to the $\mathrm{CPT}$ theorem, a permanent EDM also violates the CP symmetry. 
Since new CP violating interactions, possibly mediated by particles beyond the Standard Model, are needed to explain the observed matter-antimatter asymmetry of the Universe, 
it is conceivable that a permanent EDM is induced by the virtual effects of these particles. 
The present bound \eqref{EDMlimit} already constrains quite generically new physics (as long as it violates CP) at the TeV scale \cite{PospelovRitz2005,Engel2013}, or even at the PeV scale in some supersymmetric models \cite{McKeenPospelovRitz2013,Altmannshofer2013}. 
In addition, EDMs are sensitive probes of CP-violation in the Higgs sector \cite{Chien2015,Chen2018}. 
Therefore, experiments aiming at measuring a fundamental EDM more precisely have great discovery potentials, and it is important to measure the EDM of different systems (neutron, proton, neutral atoms and molecules, muons and other flavored particles) because they are complementary probes \cite{Jungmann2013,Chupp2014}. 

In this letter we discuss a new path to improve on 
the already ridiculously precise measurement \eqref{EDMlimit} of the neutron EDM. 
We propose to adjust the magnetic field to suppress the subtle systematic effect that was dominant in the  previous experiment \cite{Pendlebury2015},  which is possible in a bigger apparatus with a significant increase of statistical sensitivity. 

Let us first review how the neutron EDM is usually measured. 
The interaction of the neutron spin with magnetic ($\vec{B}$) and electric ($\vec{E}$) fields is given by the hamiltonian $- \mu_n \vec{\sigma} \vec{B} - d_n \vec{\sigma} \vec{E}$, where $\mu_n$ is the magnetic moment and $\vec{\sigma}$ are the Pauli matrices acting on the neutron spin states. When the electric and magnetic fields are static and aligned with an axis that we call $z$, the neutron spin precesses around $z$ at the Larmor angular frequency $\omega_n$: 
\begin{equation}
\hbar \omega_n = -\mu_n B - d_n E. 
\end{equation}
We apply the strongest possible electric field, however the electric term $d_n E$ is in any case much smaller than the magnetic term $\mu_n B$. In a $1 \, \muT$ field, the magnetic precession frequency is $\omega_n/2\pi = - 29$~Hz, whereas for an EDM of $10^{-26} \ e$cm, in a strong electric field of $20$~kV/cm, the electric precession frequency is only $5\times 10^{-8}$~Hz. 

Experiments are designed to measure precisely the Larmor frequency in the two polarities of the electric field (parallel and antiparallel to the magnetic field) and extract the EDM as 
$d_n = -\hbar \Delta \omega_n / 4E$. 
Ramsey's method of separated oscillating fields is used to measure the frequency: polarized neutrons (with spins aligned with $z$) are first subject to a transverse oscillating field close to resonance to induce a $\pi/2$ flip, then the spin is let to precess during a time $T$, then a second $\pi/2$ pulse is applied, finally the neutron spin is analyzed. 
From the asymmetry of counts in the up and down spin states one extracts the detuning between the applied frequency of the pulses and the Larmor frequency. 
The statistical precision of one measurement is given by: 
\begin{equation}
\label{fom}
\sigma(d_n) = \frac{\hbar}{2 \alpha |E| T \sqrt{N}}, 
\end{equation}
where $T$ is the time during which the spins precess in the fields, $N$ is the total number of neutron counts, and $0 \leqslant \alpha \leqslant 1$ is the polarization of the neutrons at the end of the precession period. 

Notice at this point that the statistical sensitivity is independent of the magnitude of the applied magnetic field $B_0$. 
The choice of this important parameter was governed by practical matters and not by a sensitivity optimization. 
In the first experiment with a thermal neutron beam \cite{Smith1957}, Smith, Purcell and Ramsey used a field of $25$~mT. 
In the next experiments with a cold neutron beam, the magnetic field was lowered to reach $1.7$~mT in the last experiment of this type \cite{Dress1977}. 
Then, starting from the 1980s, the technique of ultracold neutron (UCN) storage was used, in order to gain sensitivity through an increase in the precession time $T$. 
In the most recent UCN experiment \cite{Pendlebury2015}, a field of $B_0=1 \, \muT$ was applied to the neutrons stored in a cylindrical chamber of diameter 47~cm and height 12~cm. 
In order to prevent the depolarization of the neutrons during the precession time of about $T = 100$~s, the uniformity of the field in the chamber must be typically $\sigma B < 1$~nT \cite{Abel2018}. 
This requirement, which applies to the absolute uniformity and not the relative uniformity, is  better realized in lower fields. 
We will argue that we could instead use the freedom to choose the value of $B_0$ to relax considerably the systematic effects. 

One of the most dangerous systematic effect in neutron EDM experiments originates from the relativistic motional field
\begin{equation}
\label{motionalField}
\vec{B}_m = \vec{E} \times \vec{v} /c^2
\end{equation}
experienced by a particle moving at velocity $\vec{v}$ in an electric field. 
In old beam experiments, all neutrons have a velocity transverse to the fields. 
Since the electric and magnetic fields cannot be perfectly aligned, the motional field has a component parallel to the magnetic field which reverses sign at electric polarity reversal an thus it produces a false EDM. 

In modern UCN storage experiments the average velocity of the neutrons during the precession is zero. 
The influence of the motional field is therefore much reduced. 
However it does not disappear completely, we will now describe the residual systematic effect for particles trapped in a chamber in some details. 
Because the velocity changes rapidly in time, the motional field \eqref{motionalField} is now a fluctuating random field transverse to the static field $B_0 \vec{e}_z$. 
The transverse fluctuation is conveniently described by a complex perturbation, 
\begin{equation}
b(t) = (\vec{B}_m(t)+\vec{B}(t)) \cdot (\vec{e}_x + i \cdot \vec{e}_y).
\end{equation}
We will be concerned by the autocorrelation function of the perturbation 
$\langle b^*(t_1) b(t_2) \rangle$, where the brackets $\langle \cdot \rangle$ denote the ensemble average over all particles in the chamber. 
Here we assume that the motion of the particles is stationary in the statistical sense and therefore $\langle b^*(t_1) b(t_2) \rangle = \langle b^*(0) b(t_2-t_1) \rangle$. 

Any random transverse field induces a shift of the angular Larmor frequency. 
Spin relaxation theory (see \cite{Pignol2015} and references therein) allows to calculate the shift at second order in the perturbation $b$: 
\begin{equation}
\label{frequencyShift}
\delta \omega = \frac{\gamma^2}{2} \int_0^\infty {\rm Im} \left[ e^{i \omega t} \langle b^*(0) b(t) \rangle \right] dt. 
\end{equation}
In this expression, $\omega = \gamma B_0$ is the angular Larmor frequency, $\gamma$ is the gyromagnetic ratio.
If the perturbation $b$ is only due to the motional field \eqref{motionalField}, then the frequency shift is quadratic in $E$. This case was first discussed by Lamoreaux \cite{Lamoreaux1996}. 
Then Pendlebury {\it et al.} \cite{Pendlebury2004} pointed out the other  contribution $B(t) = \vec{B}(t) \cdot (\vec{e}_x + i \cdot \vec{e}_y)$ to  $b(t)$ from the motion of the particle in the static but non-uniform field, producing a term in \eqref{frequencyShift} linear in electric field which corresponds to a false EDM. 
The linear-in-electric-field frequency shift has been extensively studied theoretically \cite{Lamoreaux2005,Barabanov2006,Clayton2011,Pignol2012,Swank2012,Golub2015,Swank2016} and verified experimentally \cite{Afach2015}. 

\begin{figure*}
\includegraphics[width = \linewidth]{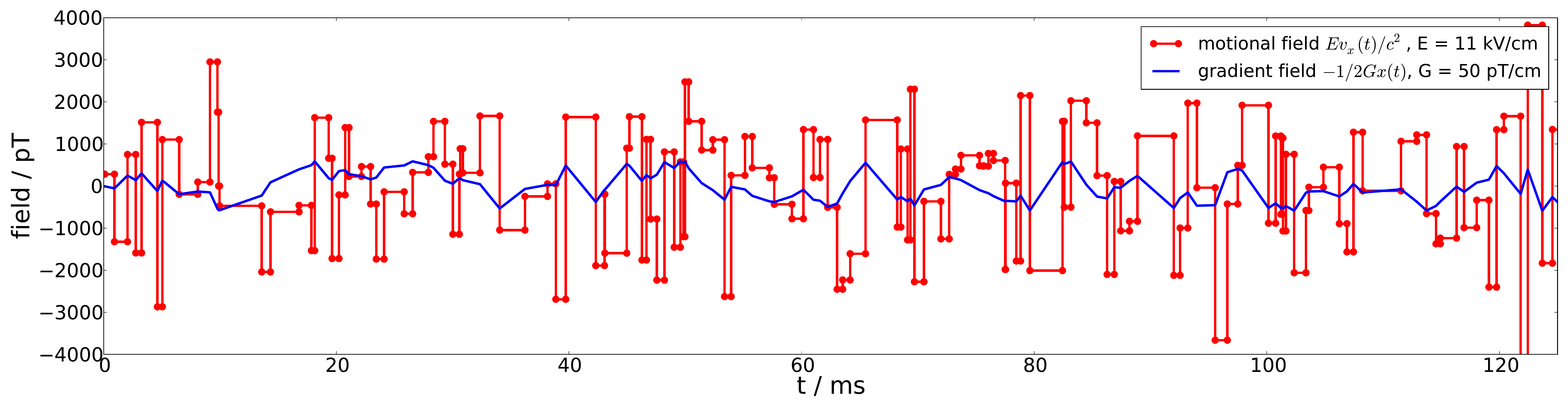}
\caption{\label{fig:motionalAndGradientFieldPlot}
MonteCarlo simulation of the transverse field seen by a mercury atom in thermal ballistic motion inside a cylinder (height $12$~cm, diameter $47$~cm) permeated with a vertical electric field of $11$~kV/cm and a linear field gradient $G_1 = 50$~pT/cm. Red lines with dots: motional field along $y$, blue line: non-uniform field along $x$. 
}
\end{figure*}

This effect affects not only the neutrons, but more importantly the atoms of the co-magnetometer. 
Indeed, in the experiment \cite{Pendlebury2015}, the magnetic field variations were monitored by measuring the Larmor frequency of polarized mercury-199 atoms precessing in the same volume and in the same time as the neutrons. 
Therefore a false EDM of the mercury $d_{\rm Hg}^{\rm false}$ translates to a false neutron EDM as $d_n^{\rm false} = |\gamma_n/\gamma_{\rm Hg}| d_{\rm Hg}^{\rm false}$ which largely dominates over the false EDM directly induced by the motional field of the neutrons. 

The co-magnetometer is essential in present and future experiments to compensate the random fluctuations of $B_0$ \emph{and} the fluctuations correlated with the electric field polarity due to leakage currents for example. 
The future experiment at Oak Ridge based on the concept proposed by Golub and Lamoreaux \cite{Golub1994} will utilize a helium-3 co-magnetometer in a UCN chamber filled with superfluid helium, while the experiments planned at TRIUMF, ILL and PSI will use a mercury co-magnetometer operating at room temperature. 
Concerning the concept in superfluid helium, a well defined strategy exists to control the false EDM of helium-3 by adjusting the temperature of the bath \cite{Barabanov2006}. 
We will now present a strategy to control the false EDM in the case of a room-temperature experiment with a mercury co-magnetometer by adjusting the magnetic field $B_0$. 

Figure \ref{fig:motionalAndGradientFieldPlot} shows a simulation of the transverse field seen by a mercury atom in the chamber of the experiment \cite{Pendlebury2015}. 
It was obtained with a dedicated Monte Carlo code \cite{TOMAt} calculating ballistic trajectories (only wall collisions, as practically verified in reality for such a rarefied gas) and randomizing the velocity at each collision with a wall at room temperature. 
We show the contributions of the motional field and of a non-uniform field, in this case a linear field gradient of the form $B_x = - G_1 x/2$ with a rather large gradient value of $G_1 = 50$~pT/cm. 
The false EDM is produced by the correlation between these two contributions. 
In fact the general expression for the false EDM derived from the linear-in-electric-field term in eq. \eqref{frequencyShift} reads: 
\begin{equation}
\label{falseEDM}
d^{\rm false}_n = \frac{\hbar |\gamma_n \gamma_{\rm Hg}|}{2 c^2} \int_0^\infty \langle B_x(0) v_x(t) + B_y(0) v_y(t) \rangle  \cos \omega t \, dt.
\end{equation}
This expression is valid for any angular frequency of the mercury spins $\omega$ and for any shape of the magnetic field. 

\begin{figure}
\includegraphics[width = \linewidth]{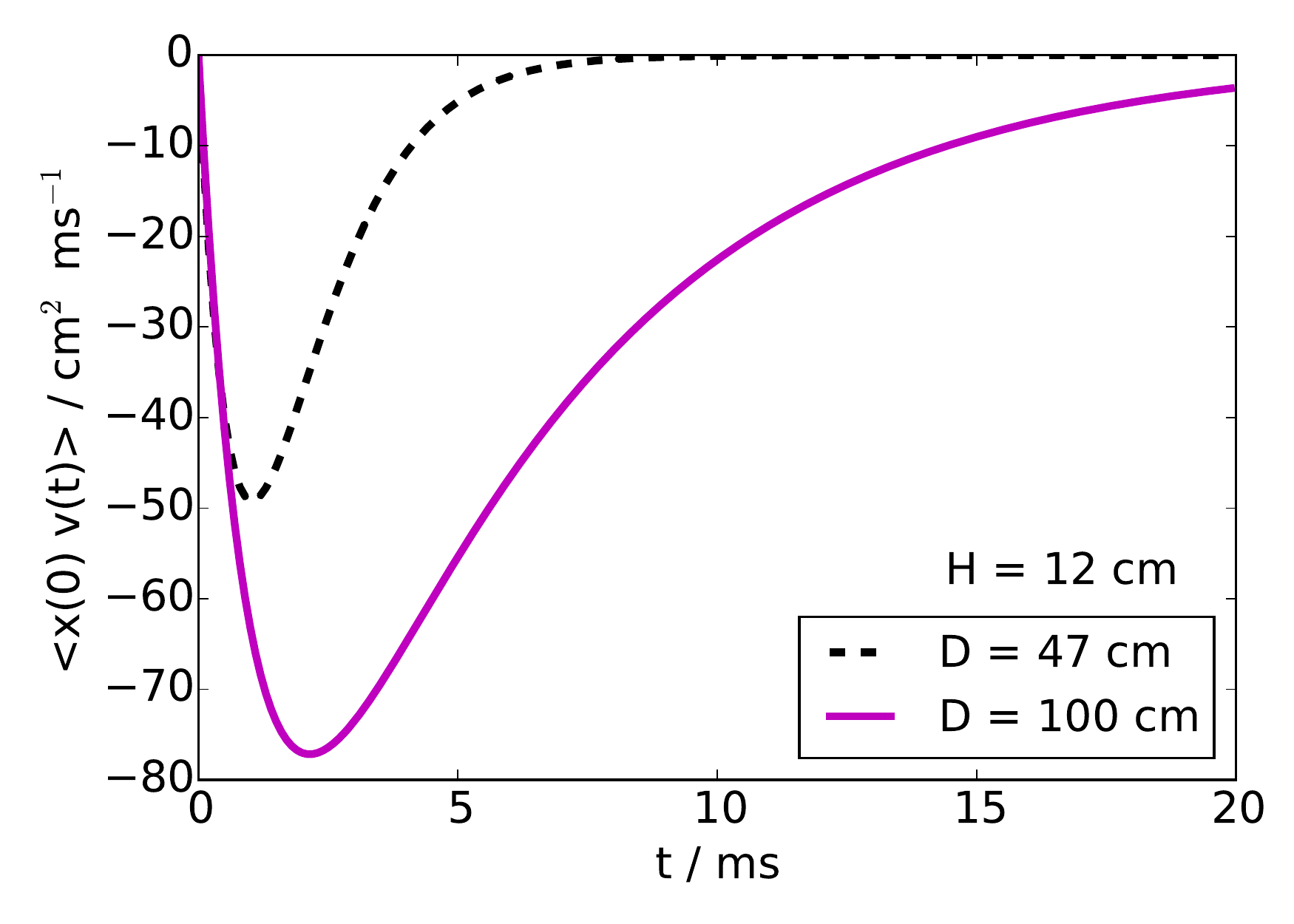}
\caption{\label{fig:correlation}
Correlation function $\langle x(0) v_x(t) \rangle$ for the random ballistic motion of mercury atoms in a cylindrical chamber of height $12$~cm and diameter $47$~cm (dashed line) or $100$~cm (plain magenta line). 
}
\end{figure}

In the case of a linear gradient, the false EDM depends on the correlation function between the position and the velocity $\langle x(0) v_x(t) \rangle$. 
Figure \ref{fig:correlation} shows the result of a numerical calculation of the position-velocity correlation function. 
The short-time behavior is independent of the size of the chamber. 
Indeed for a gas at thermal equilibrium we have
\begin{equation}
\left. \frac{d}{dt} \langle x(0) v_x(t) \rangle \right|_{t=0} = - \langle v_x^2 \rangle = -\frac{k_B T_{\rm gas}}{m} \approx -11 \ \rm{cm/ms}, 
\end{equation}
where $m$ is the mass of a mercury-199 atom, $k_B$ is Boltzmann's constant, and $T_{\rm gas}$ is the temperature (here $20 ^\circ$C). 
On the other hand the long-time behavior of the correlation function is set by the horizontal diffusion across the chamber and we expect the time constant of the decay of the correlation to scale as $D^2$. 

\begin{figure}
\includegraphics[width = \linewidth]{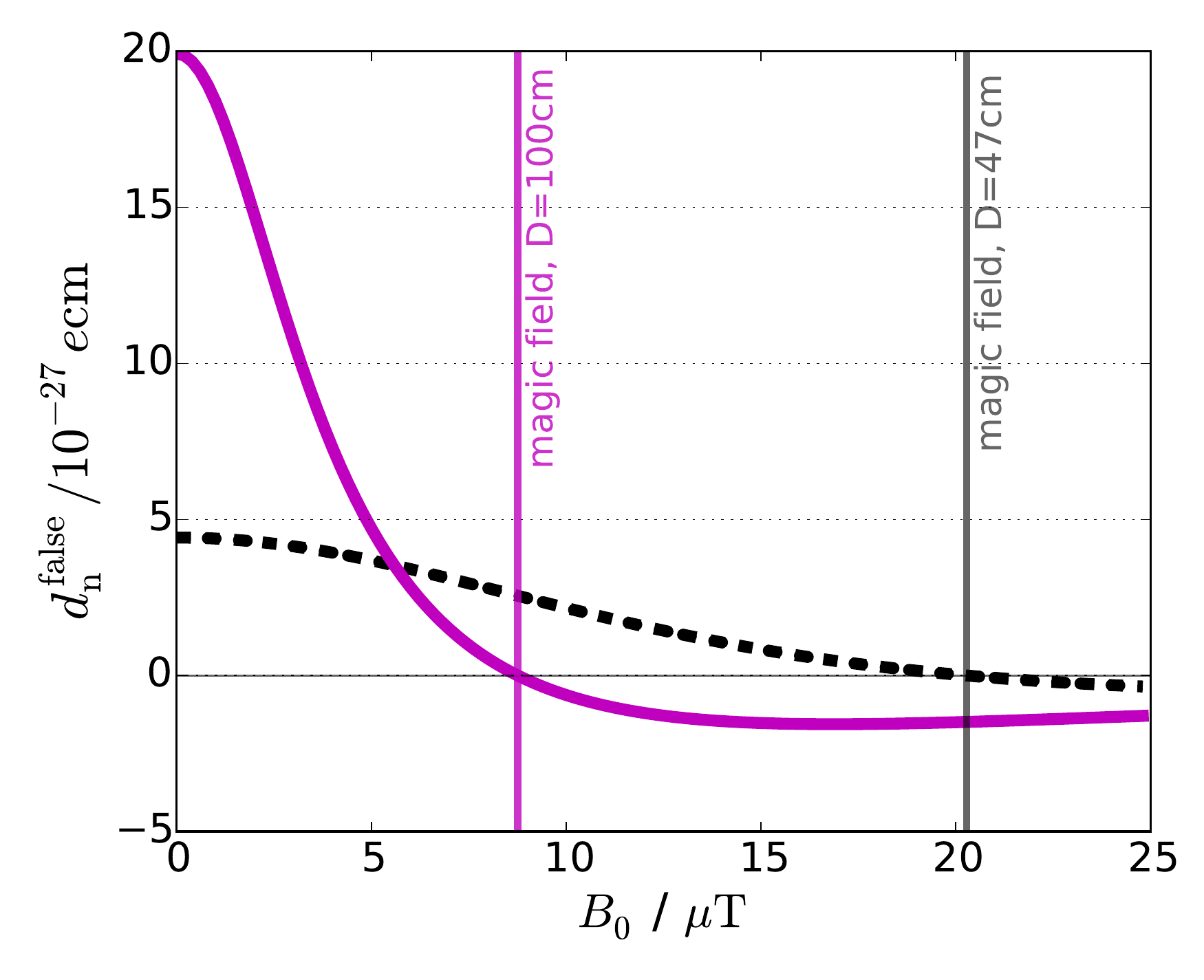}
\caption{\label{fig:dFalse}
False motional EDM $d^{\rm false}_n$ induced by a linear gradient of $G_1 = 1$~pT/cm as a function of the holding field $B_0$ in a cylindrical chamber of height $12$~cm and diameter $47$~cm (dashed line) or $100$~cm (plain magenta line). The vertical lines labeled ``magic field'' indicate the values of $B_0$ for which $d^{\rm false}_n = 0$. 
}
\end{figure}

Figure \ref{fig:dFalse} shows the false motional EDM \eqref{falseEDM} induced by a linear gradient as a function of the holding magnetic field. 
The effect is maximal at low field, when the frequency $\omega$ is slow compared to the typical decay time of the correlation functions. 
This was the case in the previous experiment \cite{Pendlebury2015} with $B_0 = 1 \, \muT$, corresponding to a frequency of $\omega = \gamma_{\rm Hg} B_0 = 2 \pi \times 7.6$~Hz. 
In the low field regime, Eq. \eqref{falseEDM} simplifies and we have an expression valid for for an arbitrary shape of the magnetic field  \cite{Pignol2012}: 
\begin{equation}
\label{dFalse_low_frequency}
d^{\rm false}_n = -\frac{\hbar |\gamma_n \gamma_{\rm Hg}|}{2 c^2} \langle x B_x + y B_y \rangle.  
\end{equation}
In this formula the ensemble average $\langle \cdot \rangle$ can be calculated as a simple volume average and no Monte Carlo simulation is required to evaluate the false EDM in the low field regime. 

As evident from Fig. \ref{fig:dFalse} and Eq. \eqref{dFalse_low_frequency}, the systematic effect seriously increases when enlarging the diameter of the chamber, at low field $B_0$. 
This is a significant limitation for future nEDM measurements, because we would like to increase the size of the chamber to store more neutrons and improve the statistical sensitivity. 
The problem can be evaded by choosing an increased value for the $B_0$ field in order to cancel the false EDM, as shown in Fig. \ref{fig:dFalse}. 
We call ``magic field'' the value satisfying $d^{\rm false}_n (B_{\rm magic}) = 0$. 
It should be noted that the magic field is weaker for larger chambers, 
which is favorable to meet the uniformity required not to depolarize the ultracold neutrons during the precession time. 
For a chamber height of $H = 12$~cm and a diameter of $D = 100$~cm, the magic field is $B_{\rm magic} = 8.8 \ \muT$. 
This field is low enough to be realized inside a large magnetic shield such as the one described in \cite{Altarev2015}. 

We run Monte Carlo simulations for various chamber sizes, 
namely $8 \, {\rm cm} \leqslant H \leqslant 16 \, {\rm cm}$ and $47 \, {\rm cm} \leqslant D \leqslant 120 \, {\rm cm}$. 
The results of these calculations can be summarized by the following formula: 
\begin{eqnarray}
\label{magic}
B_{\rm magic}^{\rm lin} & \approx & \frac{4 \langle v_x^2 \rangle}{\gamma D} \left[ 1 + 0.8 \left( \frac{H}{D}-0.2 \right) \right] \\ 
\label{magic_mercury}
& \approx & \frac{100 \, {\rm cm}}{D} \times \left( 0.84 + 1.33 \frac{H}{D} \right) \times  8.8 \, \muT. 
\end{eqnarray}
Equation \eqref{magic} can be applied to other atomic magnetometers, such as xenon for example, by taking the appropriate values for $\langle v_x^2 \rangle$ and $\gamma$. In fact the formula \eqref{magic} is valid for a gas of spins in thermal equilibrium with the walls, with purely diffuse reflection at wall collisions and in ballistic motion in a cylindrical chamber. 
Equation \eqref{magic_mercury} is specific to mercury atoms. 

At this point it is important to note that Eq. \eqref{magic} and Fig. \ref{fig:dFalse} are calculated in the case of a linear gradient only. 
More complicated modes of the non-uniformity have different values for the magic field. 
In fact, since the linear mode is the one which is easiest to measure, it might be desirable to tune the field to cancel the false EDM induced by more complicated modes. 
Let us consider the specific case of a double chamber experiment, with two vertically stacked cylinders of diameter $D = 100$~cm, height $H = 12$~cm, and a distance between the two chamber centers of $H' = 17$~cm. 
A linear mode $G_1$ can be measured from the the magnetic readings in the top and bottom chambers. 
However there could be modes of the field which are invisible in this way, we call these modes \emph{phantom fields}. 
The lowest order phantom field producing a false EDM is a combination of a linear and a cubic mode with cylindrical symmetry: 
\begin{eqnarray}
B_z & = & G_{\rm ph} \, z \left( 1 + \frac{z^2}{L^2} - \frac{3\rho^2}{2L^2} \right), \\
B_\rho & = & - G_{\rm ph} \, \rho \left( \frac{1}{2}  + \frac{3z^2}{2L^2} - \frac{3\rho^2}{8 L^2} \right), 
\end{eqnarray}
with $\rho^2 = x^2+y^2$ and 
\begin{equation}
L^2 = \frac{3}{16} D^2 - \frac{1}{4} (H^2 + H'^2). 
\end{equation}
The amplitude $G_{\rm ph}$ of the phantom mode has the dimensions of a field gradient. 
We have calculated the magic field for the phantom mode in the specific double-chamber geometry and found
\begin{equation}
B_{\rm magic}^{\rm ph} = 7.2 \, \muT, 
\end{equation}
which is about 20 \% lower than the magic field for the linear mode. 
We plan to elaborate on the implementation of the magic field concept in a large double-chamber experiment in a separate and longer article. 
The effect of even more complicated fields, in particular those generated by localized dipoles, needs to be carefully evaluated. 

Finally, let us estimate the potential sensitivity of a future nEDM experiment with such a large double-chamber. 
The experiment should use an ultracold neutron source based on superthermal production in solid deuterium, suitable to fill large volumes. 
For definiteness we consider the UCN facility at the Paul Scherrer Institute, where a density of $20$~UCN/cm$^3$ (unpolarized) has been measured in a 32~L ``standard'' bottle made out of stainless steel \cite{Bison2017}. 
This converts in an expected initial density of polarized neutrons of $10$~UCN/cm$^3$. 
A similar density has been reported for the deuterium source at Los Alamos \cite{Ito2018}. 
We consider chambers made of DLC coated electrodes (Fermi potential of $250$~neV and a loss coefficient at wall collision of $3.5 \times 10^{-4}$) as demonstrated in \cite{Atchison2005} and two insulating rings coated with deuterated polyethylene (Fermi potential of $220$~neV and a loss coefficient of $1.3 \times 10^{-4}$) as demonstrated in \cite{Brenner2015}. 
As for the initial energy spectrum of neutrons we take the one given in Fig. 32 in \cite{Bison2017}. 
With all these assumptions the number $N$ of surviving neutrons in the chambers after a storage time of $T = 180$~s can be calculated with standard formulas (namely Eqs. (7.8) and (7.9) in \cite{Golub}), we find $N = 650,000$. 
Next, assuming a final polarization of $\alpha = 0.75$ and an electric field strength of $E = 18$~kV/cm (as obtained in the previous double-chamber experiment \cite{Serebrov2015}), according to Eq. \eqref{fom} we obtain a statistical sensitivity of $\sigma d_n = 1.7 \times 10^{-25} \, e \, {\rm cm}$ for a single cycle. 
In five calendar years it is possible to produce $150,000$ cycles and the final statistical sensitivity will be $4.4 \times 10^{-28} \, e \, {\rm cm}$. 

This estimated statistical sensitivity, based on the current performance of the UCN source and achieved values for the key parameters, would represent an improvement by more than a factor 30 compared to the previous best experiment \cite{Pendlebury2015}. 
The large double-chamber concept is competitive with other proposals for a next generation nEDM measurement, such as the ongoing ambitious project at the US spallation Neutron Source based on cryogenic concept \cite{Golub1994}, or the recent concept using a pulsed beam \cite{Piegsa2013}. 

In summary, the motional field $\vec{E} \times \vec{v} /c^2$ on the co-magnetometer atoms is known to produce a false EDM when the magnetic field is not perfectly uniform. 
This constitutes the most serious systematic effect in room-temperature experiments using an atomic co-magnetometer at low field $B_0$. 
%This effect requires a control of the uniformity of the field much better than what is needed to simply prevent the depolarization of the neutrons. 
For next generation experiments, the motional false EDM would normally forbid a significant increase of the size of the chamber because the systematic effect scales with the mean square of the non-uniformity in the chamber. 
However we argued here that it is possible to tune the field $B_0$ to a ``magic'' value which cancels the false EDM. 
For a large chamber the magic field is not too high, $7.2 \, \muT$ in our specific geometry, and still allows the experiment to be placed inside a magnetically shielded room. 
This allows to improve significantly on the statistical sensitivity by using a large double-chamber, equipped with a co-magnetometer to correct for the uncorrelated and correlated field drifts, while keeping the motional false EDM in control. 
%This concept could be applied to any UCN experiment using an atomic co-magnetometer. 
At existing UCN sources, an experiment with a sensitivity of $\sigma d_n < 5 \times 10^{-28} \, e \, {\rm cm}$ is possible. 
%Should more intense sources of UCNs be operational, delivering a density of more than $1000$~UCN/cm$^3$ as claimed to be possible, a truly sensitive nEDM experiment with $\sigma d_n < 10^{-28} \, e \, {\rm cm}$ could certainly be designed by exploiting the magic field. 

%\section{acknowledgments}
\begin{acknowledgments}
I am grateful to 
G. Bison, 
B. Golub, 
P. Koss, 
D. Rebreyend, 
A. Leredde, 
S. Roccia and 
P. Schmidt-Wellenburg for valuable discussions. 
This work is supported by the European Research Council, ERC project 716651 - NEDM. 
\end{acknowledgments}

\bibliographystyle{unsrt}

\end{document}